\documentclass[aps,prl,reprint,superscriptaddress,nofootinbib]{revtex4-1}

\usepackage{amsmath,amssymb,amsfonts}
\usepackage{graphicx}
\usepackage{bm}
\usepackage{hyperref}
\usepackage{xcolor}
\usepackage{comment}

\newcommand{\mpl}{M_{\rm Pl}}

\def\mGk{\mathcal{G}_k}
\def\mPb{\mathcal{P}_{\rm B}}
\def\vk{\mathbf{k}}
\def\vq{\mathbf{q}}
\def\vx{\mathbf{x}}

\def\mPhm{P_{\rm h}^{(B)}}
\def\mPhs{P_{\rm h}^{(\zeta)}}
\def\mPb{\tilde{\mathcal{P}}_{\rm B}}

\def\d{\rm d}
\def\nn{\nonumber}
\def\mIrds{\mathcal{I}_{\rm RD, s}}
\def\mIrdm{\mathcal{I}_{\rm RD, m}}
\def\mPc{\mathcal{P}_\zeta}
\def\mrP{\mathrm{P}}
\def\Phl{P_h^{\lambda}}
\def\hkl{h_{\vk}^{\lambda}}
\def\l{\left}
\def\r{\right}
\def\hij{h_{ij}}

\def\rhob{\rho_{\rm B}}
\def\rhoc{\rho_{\rm cri}}

\def\ogw{\Omega_{\rm gw}}
\def\ogws{\Omega_{\rm gw}^{\rm scalar}}

\def\kc{k_{\rm cut}}
\def\kp{k_{\rm p}}

\def\ngw{n_{\rm gw}}
\def\hz{\mathrm{Hz}}

\def\fp{f_{\rm p}}
\def\nuv{n_{\rm UV}}
\def\nir{n_{\rm IR}}

\def\Ci{\rm Ci}
\def\Si{\rm Si}

\def\rhogw{\rho_{\rm gw}}
\def\mPh{P_h}
\def\Hz{\rm Hz}
\def\alphagw{\alpha_{\rm gw}}

\begin{document}

\title{
    Spectral Fingerprints Beyond Degeneracies in Primordial Gravitational-Wave Sources
}

\author{Subhasis Maiti}
\email{E-mail: subhashish@iitg.ac.in}
\affiliation{Department of Physics, Indian Institute of Technology, Guwahati, 
Assam, India}

\begin{abstract}
Stochastic gravitational-wave backgrounds from different primordial mechanisms can exhibit identical peak frequencies and amplitudes, leaving their origin degenerate. We show that this degeneracy can be broken by the broadband spectral response of the gravitational-wave source. By comparing scalar-induced and gauge-field-induced gravitational waves with the same primordial spectral profile and cosmological evolution, we isolate the imprint of the tensor-source dynamics. Although both mechanisms exhibit the same infrared scaling, they develop distinct spectral shapes near the peak and different ultraviolet asymptotes. The local spectral index $\ngw$ and its running $\alphagw$ further characterize this source-dependent evolution, providing a spectral fingerprint beyond the peak observables. These results show that broadband, multi-frequency observations can probe the production mechanism of a primordial gravitational-wave background even when its peak frequency and amplitude are degenerate.
\end{abstract}

\maketitle

\paragraph{\bf Introduction:}
A primordial stochastic gravitational-wave background (SGWB) provides a probe of the early Universe, carrying information about physical processes at energies and epochs inaccessible to electromagnetic observations. The detection of gravitational waves by the LIGO–Virgo Collaboration~\cite{LIGOScientific:2016aoc,LIGOScientific:2016vlm,LIGOScientific:2016emj,LIGOScientific:2016vbw,LIGOScientific:2017bnn,LIGOScientific:2016jlg} and recent evidence for a nanohertz SGWB from pulsar timing arrays~\cite{NANOGrav:2023gor,2023arXiv230616224A,Reardon:2023gzh,Zic:2023gta,Xu:2023wog} have established gravitational-wave spectroscopy as a probe of cosmology. A central challenge is to determine the physical origin of a primordial SGWB and distinguish the mechanisms that generate its tensor perturbations.

Several early-Universe processes can source gravitational waves. Enhanced scalar fluctuations generate tensor modes through nonlinear mode coupling after horizon re-entry~\cite{PhysRevD.64.123514,Di:2017ndc,Fu:2019vqc,PhysRevD.103.083510,Bhaumik:2020dor,Solbi:2021wbo,Figueroa:2021zah,Ragavendra:2020sop,Ragavendra:2023ret,Domenech:2025ffb,Domenech:2024rks,Balaji:2023ehk,Balaji:2022dbi,Domenech:2021ztg,Domenech:2020xin,Domenech:2020kqm,Domenech:2019quo,Dimastrogiovanni:2022eir,Chakraborty:2024rgl,Papanikolaou:2022chm,Maity:2024odg,Kohri:2018awv,Cai:2018dig,Cai:2019cdl,Inomata:2016rbd,Ciprini:2026pvz,Maleknejad:2016qjz}, while gauge fields source tensor perturbations through their anisotropic stresses~\cite{Sorbo:2011rz,Caprini:2014mja,Ito:2016fqp,Sharma:2019jtb,Okano:2020uyr,Maleknejad:2025clz,Maiti:2026hsn,Maiti:2026lvx,Maiti:2024nhv,Maiti:2025ijr,Maiti:2025ijr,Maiti:2025awl,Maiti:2025cbi,Maiti:2025rkn,Bhaumik:2025kuj,Ragavendra:2026fgs,Maiti:2026lvx,Bhaumik:2026rzf}. Despite their different origins, these mechanisms can produce SGWBs with similar peak frequencies and amplitudes, making peak observables alone insufficient to identify the source.

This raises the central question: if two primordial GW backgrounds are constructed to have the same peak frequency and amplitude, what information remains that can reveal their origin? In this Letter, we show that broadband spectral information can discriminate between source classes that are degenerate in peak observables, even when they share the same primordial spectral profile and cosmological evolution. For several representative profiles, we normalize the resulting SGWBs to the same peak frequency and amplitude, thereby removing the degeneracy in these observables. We then characterize the broadband spectral evolution through the local spectral index and its running. Although the resulting backgrounds can share the same peak observables and infrared scaling, their spectral evolution around the peak and ultraviolet asymptotes retain distinct signatures of the underlying tensor sources. These features reveal source-dependent information beyond the peak frequency and amplitude.

\paragraph{\bf Tensor Power Spectrum:}
Primordial anisotropic stresses source tensor perturbations~\cite{Caprini:2018mtu}. Although the propagation of tensor modes is universal, their sourcing depends on the physical nature of the primordial perturbations. The tensor perturbations $\hij(\vx,\eta)$ obey~\cite{Caprini:2018mtu,Guzzetti:2016mkm,Domenech:2021ztg}
\begin{align}\label{eq:hk_m}
h_{ij}''+2\mathcal{H}h_{ij}'-\nabla^2 h_{ij}
=
P^{mn}_{ij}
\left[
\mathcal{S}_{mn}
+\frac{2}{M_{\rm Pl}^2}T_{mn}
\right],
\end{align}
where $\mathcal{S}_{mn}$ and $T_{mn}$ denote the effective scalar~\cite{PhysRevD.64.123514,Di:2017ndc,Fu:2019vqc,PhysRevD.103.083510,Bhaumik:2020dor,Solbi:2021wbo,Figueroa:2021zah,Ragavendra:2020sop,Ragavendra:2023ret,Domenech:2025ffb,Domenech:2024rks,Balaji:2023ehk,Balaji:2022dbi,Domenech:2021ztg,Domenech:2020xin,Domenech:2020kqm,Domenech:2019quo,Dimastrogiovanni:2022eir,Papanikolaou:2022chm,Maity:2024odg,Kohri:2018awv,Cai:2018dig,Cai:2019cdl,Inomata:2016rbd,Ciprini:2026pvz} and gauge-field anisotropic-stress sources~\cite{Sorbo:2011rz,Caprini:2014mja,Ito:2016fqp,Sharma:2019jtb,Okano:2020uyr,Maiti:2025rkn,Maiti:2025awl,Maiti:2025cbi,Maiti:2025ijr,Bhaumik:2025kuj,Ragavendra:2026fgs,Maiti:2026lvx,Bhaumik:2026rzf}, respectively, and $P^{mn}_{ij}$ projects onto the transverse-traceless component. The tensor power spectrum,
\begin{align}
\Phl(k)
=
\frac{k^3}{2\pi^2}
\left\langle
\left|\hkl\right|^2
\right\rangle,
\end{align}
therefore depends not only on the primordial spectral profile but also on the momentum structure and time evolution of the tensor source. This dependence is what distinguishes the two source responses in the spectra considered below.

\paragraph{\bf Scalar- and gauge-field-induced tensor spectra:}
We consider scalar and gauge-field sources with the same dimensionless primordial spectral profile, while allowing their physical normalizations to vary independently. The resulting tensor spectra are therefore determined not only by the common primordial spectral shape, but also by the distinct source structures and their corresponding kernels.

For scalar-induced GWs during radiation domination, tensor perturbations are generated at second order by primordial curvature perturbations, with $T_{ij}=0$. The resulting tensor power spectrum is~\cite{PhysRevD.64.123514,Di:2017ndc,Fu:2019vqc,PhysRevD.103.083510,Bhaumik:2020dor,Solbi:2021wbo,Figueroa:2021zah,Ragavendra:2020sop,Ragavendra:2023ret,Domenech:2025ffb,Domenech:2024rks,Balaji:2023ehk,Balaji:2022dbi,Domenech:2021ztg,Domenech:2020xin,Domenech:2020kqm,Domenech:2019quo,Dimastrogiovanni:2022eir,Papanikolaou:2022chm,Maity:2024odg,Kohri:2018awv,Cai:2018dig,Cai:2019cdl,Inomata:2016rbd,Ciprini:2026pvz, Pi:2020otn}
\begin{align}
\label{eq:pt_scalar}
    \mPhs(k,x) & =4\int_{0}^{\infty}\d v
    \int_{|1-v|}^{1+v}\d u
    \left(\frac{4v^2-A^2}{4uv}\right)^2\nn\\
    &\times\overline{\mIrds^2(v,u,x)}
    \mPc(vk)\mPc(uk),
\end{align}
where $A=1+v^2-u^2$, $\mPc(k)$ denotes the primordial curvature power spectrum, and $v=q/k$ and $u=|\vk-\vq|/k$ are dimensionless internal momenta. The kernel $\overline{\mIrds^2}$ describes the time evolution of the scalar-induced tensor source and the tensor Green function. For modes well inside the horizon during radiation domination, $x\gg1$, its oscillation-averaged form scales as
\begin{align}
    \overline{\mIrds^2(v,u,x\gg1)}
    \simeq x^{-2}f(u,v),
\end{align}
where $f(u,v)$ contains the nontrivial momentum dependence of the radiation-era kernel (see the Supplemental Material~\ref{eq:fvux}).The tensor spectrum is consequently determined by both the primordial scalar profile and the momentum dependence of the source kernel.

\begin{table*}[t]
\centering
\caption{Primordial profiles of the fluctuations. Here, $\kp$ denotes the characteristic peak wavenumber and $A$ the corresponding amplitude. Throughout this paper, we take $A=A_\zeta=A_B=10^{-2}$ and $\kp=1.0$.}
\label{tab:spectra}
\begin{tabular}{|c| c| c| c|}
\hline\hline
Model & Parameters & BP1 & BP2 \\
\hline

Broken power law 
$\Rightarrow
A
\begin{cases}
(k/\kp)^{\nir}, & k\leq\kp,\\
(k/\kp)^{\nuv}, & k> \kp,
\end{cases}
$
&
$\nir,\;\nuv$
&
$
\nir=3,\;
\nuv=-2.5$, &$\nir=1.0;\nuv=-3.0$
\\[0.45cm]

Gaussian 
$\Rightarrow
A\exp\!\left[-\dfrac{(k/\kp-1)^2}{2\sigma^2}\right]
$
&
$\;\sigma$
&$
\sigma=0.2$ & $\sigma=0.05$
\\[0.35cm]

Log-normal 
$\Rightarrow
A\exp\!\left[-\dfrac{\ln^2(k/\kp)}{2\sigma^2}\right]
$
&
$\sigma$
&
$
\sigma=0.1$ & $\sigma=0.3$
\\[0.35cm]

Power law with cutoff 
$\Rightarrow
A\left(\frac{k}{\kp}\right)^{\nir}
\exp\!\left[-\left(\frac{k}{\kc}\right)^{2}\right]
$
&
$\nir,\;\kc$
&
$
\nir=3,\,\kc=2 \kp$ & $\nir=2.0,\,\kc=2\kp $
\\

\hline\hline
\end{tabular}
\end{table*}


We next consider tensor modes sourced by gauge-field perturbations. Here, gauge fields refer to Abelian vector fields, such as the electromagnetic or dark-photon field. Their contribution to the tensor modes arises from the corresponding anisotropic stress, which is determined by the gauge-field strength tensor $F_{\mu\nu}$. The tensor modes are sourced by the transverse-traceless component of this anisotropic stress, leading to a tensor spectrum of the form~\cite{Ragavendra:2026fgs,Atkins:2025pvg,Maiti:2026hsn}
\begin{align}
\label{eq:pt_mag}
\mPhm(k,x) =\, & 9\int_0^{\infty}\d v
\int_{|1-v|}^{1+v}\d u\,
\frac{(4v^2+A^2)(4u^2+B^2)}{8u^4v^4}
\nonumber\\
&\times\overline{\mIrdm^2(u,v,x)}
\mPb(uk)\mPb(vk),
\end{align}
where $B=1-v^2+u^2$ and
\begin{align}
\mPb(k)
=
\frac{1}{\rhoc}
\frac{\partial\rhob}{\partial\ln k}
\end{align}
where $\mPb$ is the fractional energy density per logarithemic wavenumber . The kernel $\overline{\mIrdm^2(u,v,x)}$ describes the time evolution of the anisotropic-stress source by the gauge-field and its subsequent response in the tensor sector. For a radiation-dominated background and $x\gg1$, it approaches (see the Supplemental Material for details)
\begin{align}
\overline{\mathcal{I}^2(u,v,x\gg1)}
\simeq
\frac{1}{2x^2}
\left[
\frac{\pi^2}{4}
+\left(\gamma+\ln x_i\right)^2
\right]
+\mathcal{O}(x^{-3}).
\end{align}
Both spectra involve a convolution with the corresponding primordial spectrum, but the momentum dependence and time evolution of the two kernels are different. Consequently, the same primordial profile produces different tensor spectra for the two sources. These differences determine how the same primordial spectral profile is imprinted on the tensor spectrum. After this sourcing stage, both gravitational-wave backgrounds undergo the same free propagation: sub-Hubble modes redshift as radiation, $\rhogw\propto a^{-4}$, while their physical momentum scales as $k/a$. During radiation domination, their energy density fraction is
\begin{align}\label{eq92}
\ogw(k,\eta)
=\frac{\rhogw(k,\eta)}{\rho_c(\eta)}
=\frac{1}{12}\frac{k^2\mPh}
{a^2(\eta)H^2(\eta)},
\end{align}
where $\rho_c=3H^2\mpl^2$ and $\mpl\simeq2.43\times10^{18},\mathrm{GeV}$. The present-day spectrum then follows from radiation-like redshifting,
\begin{align}
\ogw(k)h^2
\simeq
\left(\frac{g_{*s}}{g_{*s,\rm eq}}\right)^{1/3}
\Omega_Rh^2\,\ogw(k,\eta),
\end{align}
where $\Omega_Rh^2\simeq4.3\times10^{-5}$ and
$g_{*s,\rm eq}\simeq g_{*s}\simeq3.35$ are the effective entropy degrees of freedom at matter-radiation equality and today, respectively.

\paragraph{\bf Spectral fingerprints of the tensor source:}

To isolate the information carried by the GW spectrum about its physical origin, we compare scalar-induced and gauge-field-induced gravitational waves generated from the same dimensionless primordial spectral profile,
\begin{align}
\mathcal{P}_{\zeta}(k)=A_{\zeta}S(k/k_p),
\qquad
\mathcal{P}_{B}(k)=A_BS(k/k_p),
\end{align}
where the source amplitudes $A_{\zeta}$ and $A_B$ are independent, while the shape function $S(k/\kp)$ and the cosmological evolution are held fixed. This construction removes differences arising from the primordial spectral shape or the background expansion and isolates the dependence on the tensor-source kernels.

For a representative broken-power-law spectrum,
\begin{align}
S(k/\kp)=
\begin{cases}
(k/\kp)^{\nir}, & k\leq \kp,\\
(k/\kp)^{\nuv}, & k>\kp ,
\end{cases}
\end{align}
the infrared behavior of the induced GW spectrum is
\begin{align}
\ogw(k\ll \kp)\propto
\begin{cases}
k^{2n_{\rm IR}},&n_{\rm IR}\leq3/2,\\
k^3,&n_{\rm IR}>3/2 .
\end{cases}
\end{align}
Thus, in the deep infrared, the spectral behavior is controlled predominantly by the primordial profile and is insensitive to the detailed nature of the tensor source. For the range $n_{\rm IR}>3/2$ considered below, both mechanisms consequently approach the same universal $k^3$ scaling.

The source dependence becomes explicit in the ultraviolet. For scalar-induced gravitational waves,
\begin{align}
\ogw^{(\zeta)}(k\gg k_p)
\propto k^{n_{\rm GW}^{\rm UV}},
\end{align}
with
\begin{align}
\ngw^{\rm UV}
=
\begin{cases}
2n_{\rm UV},& \nuv\geq -4,\\
n_{\rm UV}-4,& n_{\rm UV}<-4,
\end{cases}
\end{align}
whereas gauge-field-induced gravitational waves obey
\begin{align}
\ogw^{(B)}(k\gg k_p)
\propto k^{n_{\rm UV}}.
\end{align}
The ultraviolet behavior is more sensitive to the structure of the tensor source. For the two sourcing mechanisms considered here, the corresponding kernel integrals weight the primordial spectrum differently at large wavenumbers. As a result, the high-frequency behavior need not follow the same power law even when the underlying dimensionless spectral profile is identical. The corresponding asymptotic behaviors are obtained below and can be directly compared with the numerical spectra.

To remove the trivial dependence on the overall normalization, we compare the peak-normalized spectra,
\begin{align}
\widehat{\Omega}_{\rm gw}^{X}(f)
=
\frac{\Omega_{\rm gw}^{X}(f)}
{\Omega_{\rm gw}^{X}(f_p)},
\qquad X=\zeta,B ,
\end{align}
and express the frequency dependence through $f/\fp$. After this normalization, the peak amplitude and frequency no longer distinguish the two spectra. Since the primordial profile and cosmological evolution are kept fixed, the remaining differences mainly reflect the different source kernels and their transfer functions.

\begin{figure}
    \centering
    \includegraphics[width=1.0\linewidth]{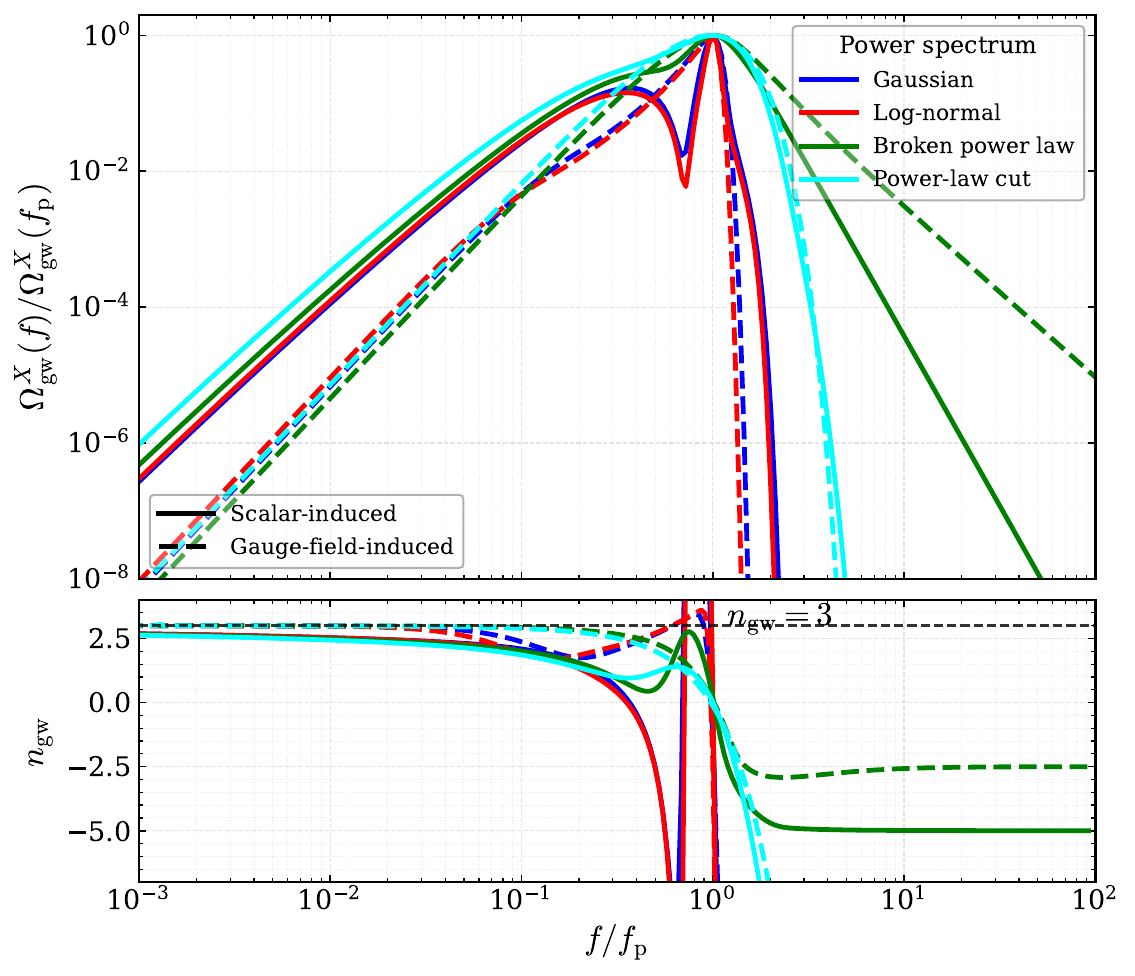}
   \caption{Top: Normalized GW spectra for the primordial profiles in Table~\ref{tab:spectra} as functions of $f/\fp$. Solid (dashed) curves show scalar-induced [gauge-field-induced] GWs for BP1. Bottom: Corresponding GW spectral indices. After aligning the peak frequency and amplitude, scalar and gauge-field sources retain distinct spectral slopes and peak morphologies, providing a source-sensitive spectral signature.}
    \label{fig:gws_nor}
\end{figure}

\begin{figure}
    \centering
    \includegraphics[width=1.0\linewidth]{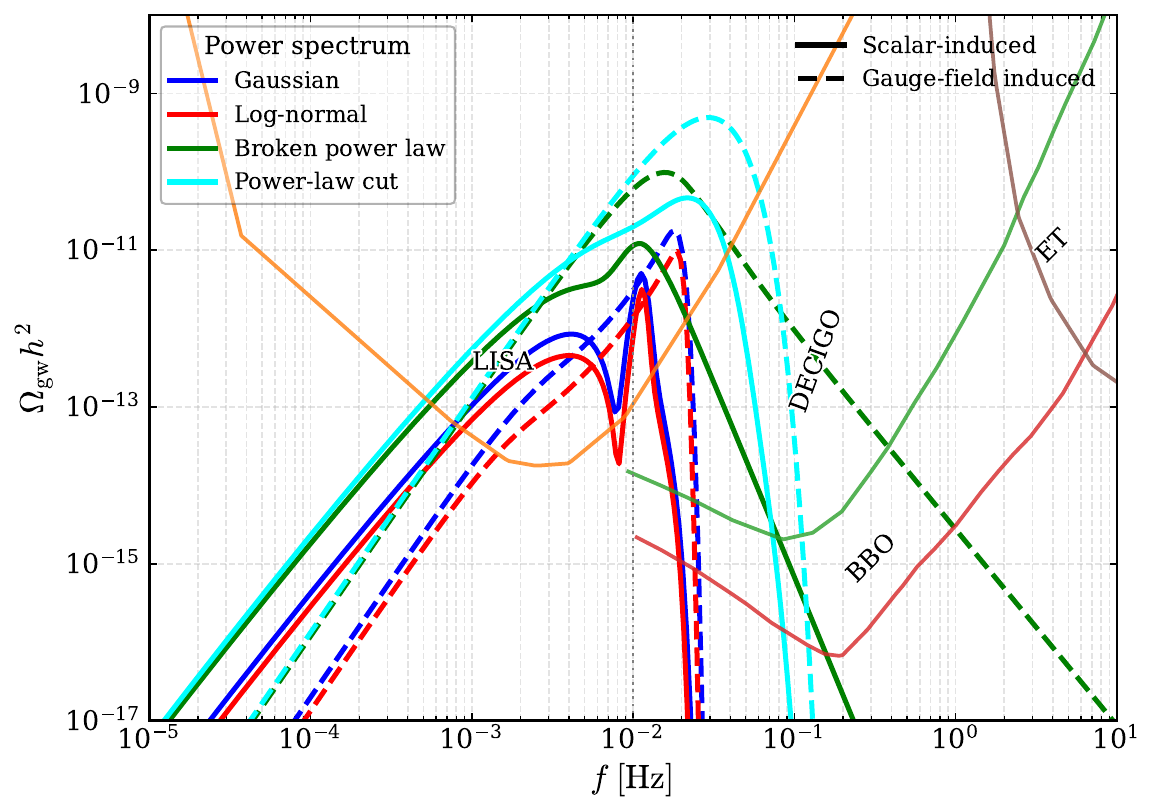}
    \caption{
Present-day GW spectra, $\ogw h^2$, as functions of the observable frequency $f$ (Hz), for the primordial spectral profiles listed in Table~\ref{tab:spectra} for BP1. The primordial peak frequency is fixed to $\fp=10^{-2}\,\hz$. Solid and dashed curves denote scalar- and gauge-field-induced GWs, respectively. The distinct spectral shapes persist after redshifting to the present epoch and provide a direct signature of the primordial source.}
    \label{fig:gws_present}
\end{figure}

Figure~\ref{fig:gws_nor} shows the normalized spectra for the four primordial spectral profiles listed in Table~\ref{tab:spectra} for BP1. Although the scalar-induced and gauge-field-induced backgrounds have identical peak frequencies and amplitudes, their broadband spectra exhibit distinct shapes. For all benchmark profiles, $\nir>3/2$, and both spectra approach the common infrared behavior $\ogw(f\ll f_p)\propto f^3$~\cite{Cai:2019cdl}. The lower panels show the corresponding local spectral index $\ngw(f)$, which removes the overall normalization and directly characterizes the frequency dependence of the spectrum. In the asymptotic infrared regime, $f\ll\fp$, all benchmark cases approach the universal value $\ngw\simeq3$, independently of the production mechanism. The source dependence instead emerges as the peak is approached. In the intermediate region, $10^{-2}f_p\lesssim f\lesssim f_p$, the scalar-induced spectra evolve more rapidly toward the peak, whereas the gauge-field-induced spectra show a smoother evolution. Above the peak $(f>\fp)$, the two mechanisms approach different asymptotic slopes. Thus, even for identical primordial spectral profiles, the broadband spectrum retains information about the tensor-source dynamics.

For the representative broken-power-law profile with $\nir=3.0,\&,\nuv=-2.5$, we find
\begin{align}
\ogw^{(\zeta)}(f>\fp)&\propto f^{-5},
\qquad
\ogw^{(B)}(f>\fp)\propto f^{-2.5}.
\end{align}
These numerical asymptotes agree with the analytical predictions (see Figure.\ref{fig:gws_nor}). Since the same primordial shape $S(k/\kp)$ and cosmological evolution are used in both calculations, the difference in the high-frequency behavior directly reflects the different tensor-source responses.

The detectability of the resulting background depends on both the source amplitude and the peak frequency. As a representative case, we take $A_{\zeta}=A_B=10^{-2}$ and $\fp=10^{-2},\Hz$ for both mechanisms. In Figure.\ref{fig:gws_present}, we plot the present-day GW spectra for these benchmark (BP1) points listed in Tab.\ref{tab:spectra} with the future sensitivity curves. The solid and dashed curves denote the scalar- and gauge-field-induced backgrounds, respectively. Despite having identical peak observables, the two spectra remain distinguishable across the sensitivity ranges of LISA~\cite{Amaro-Seoane:2012aqc,Barausse:2020rsu}, DECIGO~\cite{Seto:2001qf,Kawamura:2011zz,Suemasa:2017ppd}, BBO~\cite{Crowder:2005nr,Corbin:2005ny,Baker:2019pnp}, and ET~\cite{Punturo:2010zz,Sathyaprakash:2012jk}. This illustrates that identifying the production mechanism requires broadband spectral information rather than the peak amplitude and frequency alone, and motivates combining measurements across multiple frequency bands.

The comparison above isolates the source response by fixing the primordial spectral profile. In a realistic setting, however, the primordial spectra need not be identical, so both $\ogw$ and $\ngw$ contain information about the primordial spectral shape as well as the production mechanism. To characterize the spectral evolution beyond the local slope, we consider the running of the GW spectral index,
\begin{align}
\alphagw(f)=\frac{d\,\ngw}{{d}\,\ln f}
=\frac{d^2\,\ln\ogw}{{d}(\ln f)^2}.
\end{align}
Schematically, the GW spectrum can be written as
\begin{align}
\ogw^{(X)}\sim A^2\mathcal{S}^{(X)}(f)\mathcal{K}^{(X)}(f,\eta),
\end{align}
where $\mathcal{S}^{(X)}$ describes the primordial spectral profile and $\mathcal{K}^{(X)}$ the response of the tensor source. For a frequency-independent $A$,
\begin{align}
\alphagw=
\frac{d^2\ln\mathcal{S}(f)}{d(\ln(f))^2}
+\frac{d^2\ln\mathcal{K}(f)}{d(\ln(f))^2}.
\end{align}
Thus, $\alphagw$ is not a source-only quantity: it receives contributions from both the primordial spectral structure and the source response. Its advantage is that it probes the curvature of the spectrum in logarithmic frequency and therefore emphasizes regions where the spectral slope changes. Not only have we seen at the asymptotic limit, induced GWs behave well, but near the peak, depending on the nature of the source, it deviates from its usual scaling behaviour. As the behaviour of the sources is smooth over the scale, the deviation from the usual scaling of the GW spectrum comes due to the response term.

\begin{figure}
    \centering
    \includegraphics[width=1.0\linewidth]{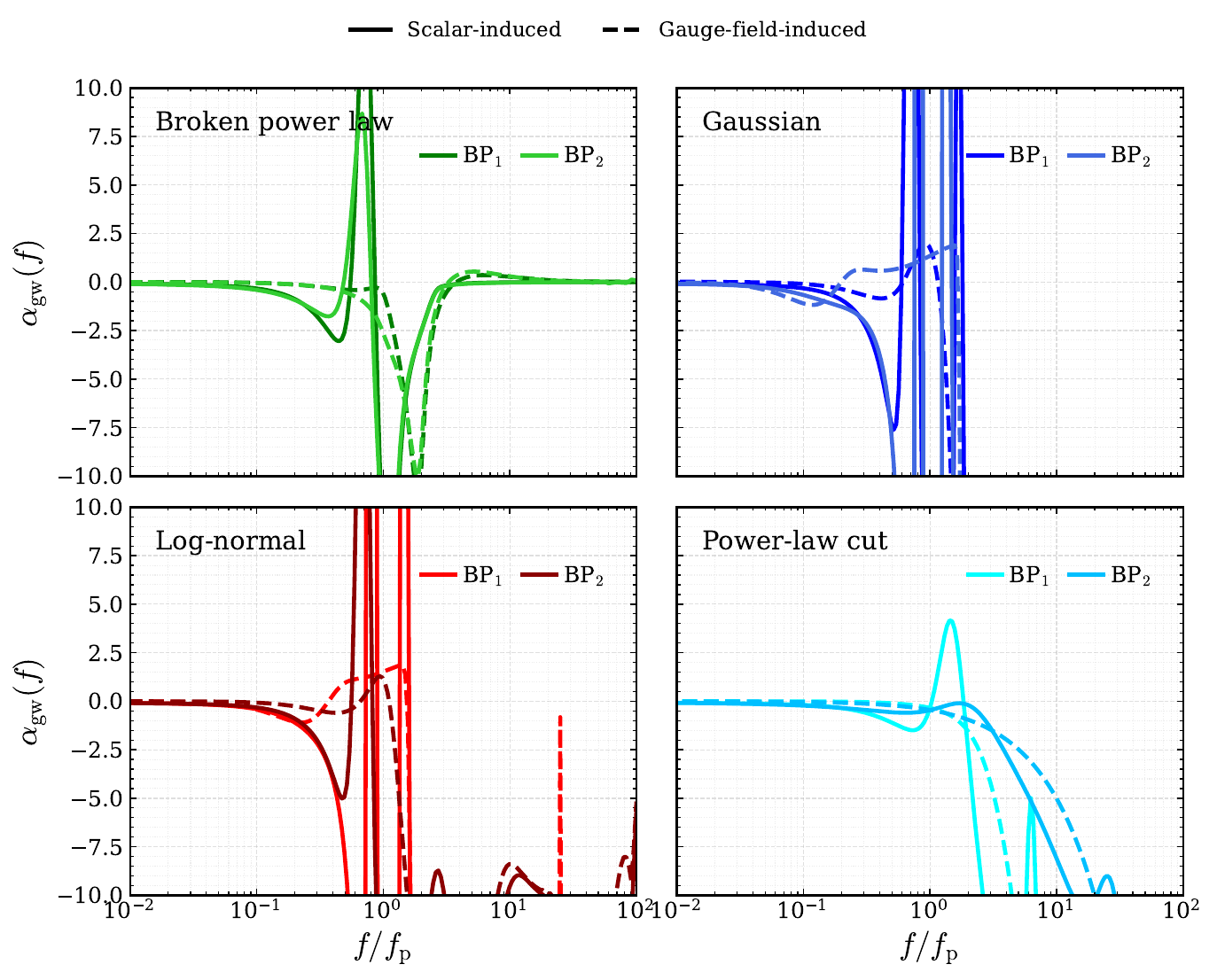}
    \caption{The running of the GW spectral index, $\alphagw$, as a function of frequency $f/\fp$ for the benchmark points listed in Table~\ref{tab:spectra}.}
    \label{fig:spec_index}
\end{figure}

Figure~\ref{fig:spec_index} shows $\alphagw$ for the four primordial profiles in Table~\ref{tab:spectra}, considering two benchmark points for each case. For a fixed primordial profile, the scalar-induced and gauge-field-induced backgrounds exhibit distinct $\alphagw(f)$ patterns even when their peak frequencies and amplitudes are identical. For the broken-power-law profile, the scalar-induced background develops a sign-changing structure around the peak, with a positive extremum for $f<\fp$ and a negative extremum for $f>\fp$, followed by $\alphagw\rightarrow0$ in both asymptotic regimes. In contrast, the gauge-field-induced background exhibits a single dominant feature near $\fp$. Similar source-dependent differences persist for the other primordial profiles, although their detailed $\alphagw(f)$ structures vary with the initial spectral shape. The overall $\alphagw(f)$ behavior differs between the two source classes, while its detailed shape also depends on the primordial profile. Thus, $\alphagw$ is not determined by the source alone. However, for a fixed primordial profile, the difference between the two source responses remains clear.

Conversely, $\alphagw$ also retains sensitivity to the primordial spectral profile. Focusing on the scalar-induced backgrounds (solid curves in Figure~\ref{fig:spec_index}), the four initial profiles produce qualitatively distinct $\alphagw$ signatures. The broken-power-law case exhibits the characteristic pair of sharp extrema around the peak, whereas the Gaussian and log-normal profiles give more similar structures. For $f<\fp$, both develop a pronounced dip-like feature followed by oscillatory behavior, reflecting the similarity of their primordial spectral profiles (see Fig.~\ref{fig:gws_nor}). The power-law-with-exponential-cut profile instead produces a qualitatively different pattern. The gauge-field-induced backgrounds show the same dependence on the primordial profile, but with source-specific $\alphagw(f)$ features. Hence, $\alphagw$ encodes both the evolution of the primordial spectral shape and the response of the production mechanism. Since $\alphagw=0$ for an exact power-law GW spectrum, its nontrivial structure is concentrated in the frequency ranges where the spectral slope evolves, particularly around the peak. The combined information in $\ogw$, $\ngw$, and $\alphagw$ therefore provides complementary probes of the primordial spectrum and tensor-source dynamics, enabling a broadband characterization of the SGWB even when its peak frequency and amplitude are degenerate.

\paragraph{\bf Discussion and Conclusions:}

Identifying the origin of a primordial stochastic gravitational-wave background (SGWB) requires information beyond its peak frequency and amplitude. Here, we isolate the imprint of the tensor source where we consider scalar- and gauge-field-induced gravitational waves with the same primordial spectral profile and cosmological evolution, while independently varying their source amplitudes. Although the resulting tensor perturbations obey the same propagation equation, their source terms and response kernels differ, leaving distinct imprints on the broadband spectrum.

For identical primordial spectral shapes and matched peak frequency and amplitude, the two mechanisms exhibit the same infrared behavior, $\ogw(f\ll f_p)\propto f^3$ for $n_{\rm IR}>3/2$, but develop different spectral evolution toward and beyond the peak. The scalar-induced background evolves more sharply around the peak, leading to stronger variations in $\ngw$ and $\alphagw$, whereas the gauge-field-induced spectrum evolves more smoothly. At high frequencies, the two backgrounds approach different ultraviolet power laws, in agreement with the analytical predictions. Since the primordial profile and cosmological evolution are fixed in this comparison, these differences directly trace the distinct tensor-source responses.

The running of the GW spectral index,
$\alphagw$, provides a complementary characterization of this spectral evolution. It probes the frequency dependence of $\ngw$ and therefore emphasizes regions where the spectral slope changes, particularly around the peak. Its detailed behavior depends on both the primordial spectral profile and the source response, but the characteristic patterns remain distinct between the scalar-induced and gauge-field-induced cases. Thus, $\alphagw$ provides information about the production mechanism that is not contained in the peak observables or in the asymptotic spectral indices alone, while simultaneously retaining sensitivity to the primordial spectral structure.

The combined frequency dependence of $\ogw$, $\ngw$, and $\alphagw$ therefore provides a broadband characterization of the SGWB beyond its peak frequency and amplitude. Measurements spanning the transition region around the peak and the ultraviolet tail can probe both the primordial spectral structure and the dynamics of the tensor source. This offers a spectral fingerprint for distinguishing different GW production mechanisms even when their conventional peak observables are degenerate.
\paragraph{Acknowledgments:}
SM acknowledges the high-energy physics division of IIT Guwahati for fruitful discussions, and is also thankful to Prof. Debaprasad Maity for many useful discussions.

\subsection{Supplemental Material}

\paragraph{\bf Tensor Power Spectrum:}
Primordial anisotropic stresses can source tensor perturbations. The tensor perturbations $\hij$ satisfy the equation of motion~\cite{Caprini:2018mtu}
\begin{align}
h_{ij}''+2\mathcal{H}h_{ij}'-\nabla^2 h_{ij}
=
P^{mn}_{ij}
\left[
\mathcal{S}_{mn}
+\frac{2}{M_{\rm Pl}^2}T_{mn}
\right],
\label{eq:hk_master}
\end{align}
where $P_{ij}^{mn}=\mrP^m_i\mrP^n_j-\frac{1}{2}\mrP^{mn}\mrP_{ij}$, with $\mrP_{ij}=\delta_{ij}-\partial_i\partial_j/\Delta$~\cite{Acquaviva:2002ud,Baumann:2007zm}. Here, $\mathcal{S}_{mn}$ represents the effective source term generated by quadratic scalar perturbations~\cite{PhysRevD.64.123514,Di:2017ndc,Fu:2019vqc,PhysRevD.103.083510,Bhaumik:2020dor,Solbi:2021wbo,Figueroa:2021zah,Domenech:2020kqm,Cai:2018dig,Domenech:2025ffb,Domenech:2024rks,Domenech:2021ztg,Domenech:2020xin,Domenech:2020kqm,Domenech:2019quo,Balaji:2023ehk,Balaji:2022dbi, Inui:2024fgk}, whereas $T_{mn}$ denotes the anisotropic stress contribution from gauge-field perturbations~\cite{Sorbo:2011rz,Caprini:2014mja,Ito:2016fqp,Sharma:2019jtb,Okano:2020uyr,Maiti:2025rkn,Maiti:2025awl,Maiti:2025cbi,Maiti:2025ijr}.

In Fourier space, the tensor perturbations can be decomposed as
\begin{align}
    \hij(\vx,\eta)=\int \frac{d^3\vk}{(2\pi)^{3/2}} e^{i\vk\cdot\vx}\Big[ h_{\vk}^{+}(\eta)e^{+}_{ij}(\vk)+h^{\times}_{\vk}(\eta)e^{\times}_{ij}(\vk)\Big].
\end{align}
The polarization tensors $e^{\lambda}_{ij}(\vk)$ ($\lambda=\{+,\times\}$) satisfy the relations
$k^ie_{ij}^{\lambda}=0$,
$e_{ii}^{\lambda}=0$,
and
$e_{ij}^{\lambda}e_{ij}^{\lambda'}=\delta^{\lambda\lambda'}$
~\cite{Caprini:2018mtu, Maiti:2026hsn}. Since the primordial sources considered in this work do not possess net helicity, the two polarization modes evolve in an identical manner. The tensor power spectrum for each polarization is therefore given by
\begin{equation}
\Phl(k)=\frac{k^3}{2\pi^2}\l|\hkl\r|^2 .
\end{equation}

Throughout this work, scalar-induced and gauge-field-induced tensor perturbations are assumed to be statistically independent. Correlations between them may arise if curvature perturbations are generated by gauge-field perturbations; however, such contributions enter the GW spectrum only at higher order and are neglected here~\cite{Maiti:2026hsn}.

\paragraph{\bf Scalar-induced tensor power spectrum:}
We first consider the GW background generated by primordial curvature perturbations during radiation domination. In this case, the source term in Eq.~(\ref{eq:hk_master}) is completely determined by the effective scalar source $\mathcal{S}_{ij}$, with $T_{ij}=0$. Following Refs.~\cite{PhysRevD.64.123514,Di:2017ndc,Domenech:2020kqm}, the corresponding tensor power spectrum is
\begin{align}\label{eq:ph_scalar}
    \mPhs(k,x) &=4\int_{0}^{\infty}\d v\int_{|1-v|}^{1+v}\d u \left(\frac{4v^2-A^2}{4u v}\right)^2\nn\\
    &\times \overline{\mIrds^2(v,u,x)}\mPc(vk)\mPc(uk)\,.
\end{align}
Here, $A=1+v^2-u^2$, $\mPc(k)$ denotes the primordial curvature power spectrum, and $x=k\eta$.

The kernel $\mIrds(v,u,x)$ describes the time evolution of the tensor modes and is defined as~\cite{Papanikolaou:2022chm}
\begin{align}\label{eq:Ird_scalar}
    \mIrds(v,u,x)=\int_{x_i}^{x}\d x_1 \tilde{\mGk}(x,x_1) f(v,u,x_1)\,,
\end{align}
where $f(v,u,x_1)$ contains the information about the scalar source contribution and is given by~\cite{PhysRevD.64.123514, Di:2017ndc, Fu:2019vqc, PhysRevD.103.083510, Bhaumik:2020dor, Solbi:2021wbo, Figueroa:2021zah, Ragavendra:2020sop, Ragavendra:2023ret, Domenech:2025ffb,Domenech:2024rks,Balaji:2023ehk,Balaji:2022dbi,Domenech:2021ztg,Domenech:2020xin,Domenech:2020kqm,Domenech:2019quo,Dimastrogiovanni:2022eir,Papanikolaou:2022chm, Maity:2024odg, Kohri:2018awv,Cai:2018dig, Cai:2019cdl}
\begin{align}\label{eq:fvux}
    &f(v,u,x_1) =\frac{4}{3}\Phi_{\vq}(vx_1)\Phi_{\vk-\vq}(ux_1)\nn\\
   & +\frac{4 x_1}{9}\Big(\partial_{\eta_1}\Phi_{\vq}(vx_1)\Phi_{\vk-\vq}(ux_1)+\Phi_{\vq}(vx_1)
    \partial_{\eta_1}\Phi_{\vk-\vq}(ux_1)\Big)\nn\\
   & +\frac{4 x_1^2}{9}\partial_{\eta_1}\Phi_{\vq}(vx_1)\partial_{\eta_1}\Phi_{\vk-\vq}(ux_1)\,.
\end{align}

In the subhorizon regime, $x\gg1$, the radiation-era kernel takes the asymptotic form~\cite{PhysRevD.64.123514, Di:2017ndc, Fu:2019vqc, PhysRevD.103.083510, Bhaumik:2020dor, Solbi:2021wbo, Figueroa:2021zah, Ragavendra:2020sop, Ragavendra:2023ret, Domenech:2025ffb,Domenech:2024rks,Balaji:2023ehk,Balaji:2022dbi,Domenech:2021ztg,Domenech:2020xin,Domenech:2020kqm,Domenech:2019quo,Dimastrogiovanni:2022eir,Papanikolaou:2022chm, Maity:2024odg, Kohri:2018awv,Cai:2018dig, Cai:2019cdl,Inomata:2016rbd}
\begin{align}
    \overline{\mIrds^2(v,u,x\rightarrow\infty)}\simeq \frac{1}{2x^2}\left( \frac{3(v^2+u^2-3)}{4v^3u^3}\right)^2\nn\\
    \times \Bigg( \left( -4vu+(v^2+u^2-3)\log\left| \frac{3-(v+u)^2}{3-(v-u)^2}\right|\right)^2\nn\\
    +\pi^2(v^2+u^2-3)^2\Theta(v+u-\sqrt{3})\Bigg)\,.
\end{align}
The dimensionless momentum variables are defined as $v=q/k$ and $u=|\vk-\vq|/k$.

\paragraph{\bf Gauge-field-induced tensor power spectrum:}
For GWs induced by gauge fields, neglecting higher-order contributions, the source term in Eq.~(\ref{eq:hk_master}) is determined solely by the anisotropic stress $T_{ij}$. The resulting tensor power spectrum is given by~\cite{Maiti:2026hsn, Ragavendra:2026fgs, Atkins:2025pvg}
\begin{align}
\label{eq:ph_mag}
\mPhm(k,x) =\, & 9\int_0^{\infty}\d v\int_{|1-v|}^{1+v}\d u\, \frac{(4v^2+A^2)(4u^2+B^2)}{8 u^4u^4}\nn\\
&\times\overline{\mIrdm^2(u,v,x)}\mPb(uk)\mPb(vk)
\end{align}
where $B=1-v^2+u^2$ and
$\mPb(k)=\frac{1}{\rhoc}\frac{\partial \rhob}{\partial \ln k}$.

The kernel $\overline{\mIrdm^2(u,v,x)}$ describes the time evolution associated with the gauge-field source and can be obtained from
\begin{align}\label{eq:Irdm}
    \mIrdm(u,v,x)=\frac{1}{x}\int_{x_i}^{x}\d x_1 \frac{\sin(x-x_1)}{x_1}
\end{align}
Performing the integral gives
\begin{align}
    \mIrdm(u,v,x) =\frac{1}{x}\{\Ci(x_1)\sin(x)-\Si(x_1)\cos(x)\}
\end{align}
where $\Ci(x)=\int \frac{\cos(x)}{x}\,\d x$ and $\Si(x)=\int \frac{\sin(x)}{x}\,\d x$. Since we are interested in modes that are deep inside the horizon during radiation domination, the integration limits in Eq.\eqref{eq:Irdm} satisfy $x\gg 1$. Moreover, when the source becomes active, all relevant modes are outside the horizon, implying $x_i\ll 1$. In the sub-Hubble limit, after averaging over oscillations, we obtain
\begin{align}
   \overline{\mIrdm^2(u,v,x\gg 1)}\simeq \frac{1}{2 x^2}\l\{ \frac{\pi^2}{4}+ (\gamma+\ln(x_i))^2 \r\}+\mathcal{O}(x^{-3})\,, 
\end{align}
where we have used
\begin{align}
    \Ci(x\gg 1)\simeq 0;\,\, \Si(x\gg 1)\simeq \frac{\pi}{2}\nn\\
    \Si(x_i\ll1)\simeq x_i;\,\, \Ci(x_i\ll1)\simeq \gamma+\ln(x_i)
\end{align}

\subsection{Extended Results for Different Benchmark Points}

\begin{figure}
    \centering
    \includegraphics[width=1.0\linewidth]{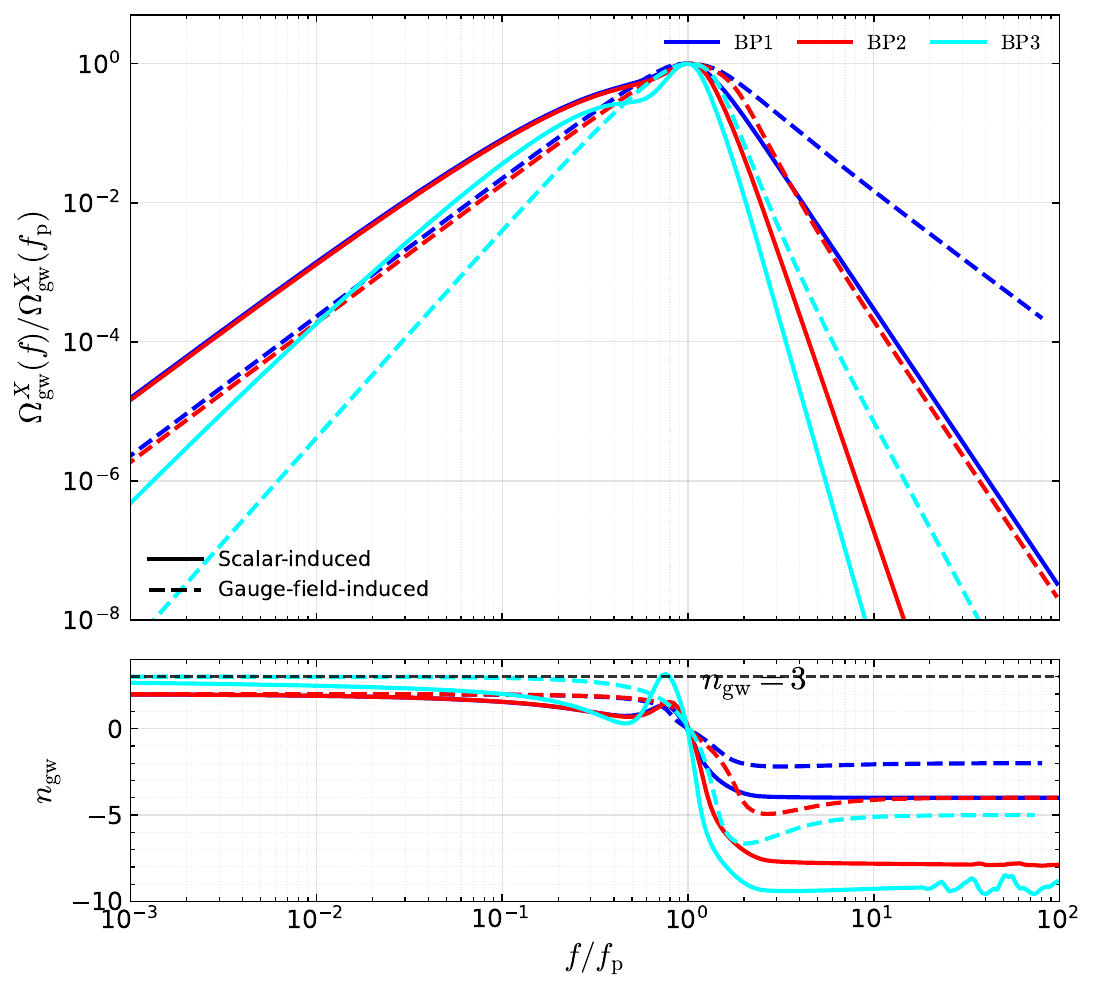}
    \caption{Normalized GW spectra as a function of frequency normalized by the peak frequency, $f/\fp$, for the benchmark points listed in Tab.~\ref{tab:bp_2}. Different colors correspond to different benchmark points, while solid and dashed lines represent the scalar- and gauge-field-induced contributions, respectively.}
    \label{fig:sgw_bp}
\end{figure}

To assess the robustness of the spectral differences discussed in the main text, we consider three representative benchmark points (BPs) for the broken-power-law initial spectrum, listed in Tab.\ref{tab:bp_2}. In Fig.\ref{fig:sgw_bp}, we show the GW spectra normalized by their respective peak amplitudes, $\ogw^{X}(f)/\ogw^{X}(\fp)$, as a function of the normalized frequency $f/\fp$. The solid and dashed curves denote the scalar-induced and gauge-field-induced contributions, respectively, while different colors correspond to the three BPs. This normalization removes the dependence on the overall amplitude and highlights the intrinsic differences in the spectral shapes.

\begin{table}[t]
\centering
\caption{Benchmark parameter space considered for the broken-power-law initial spectrum. Here, $\nir$ and $\nuv$ denote the infrared and ultraviolet spectral indices, respectively, while $\kp$ characterizes the peak wavenumber.}
\label{tab:bp_2}
\begin{tabular}{c c c c}
\hline\hline
Benchmark Point & $\nir$ & $\nuv$ & $\kp$ \\
\hline
BP1 & 1.0 & -2.0 & $1.0$\\
BP2 & $1.0$ & $-4.0$ & $1.0$\\
BP3 & $2.0$ & $-5.0$ & $1.0$\\
\hline
\end{tabular}
\end{table}

The two source channels exhibit similar behavior in the deep infrared but significantly different ultraviolet tails. For BP1, with $\nir=1.0$ and $\nuv=-2.0$, both spectra scale as $\ogws(f\ll\fp)\propto f^2$. This common infrared behavior follows from the fact that both channels probe the same large-scale branch of the primordial spectrum. In contrast, their UV tails differ, with $\ogws(f>\fp)\propto f^{-4}$ for the scalar-induced contribution and $\ogw^{(\rm B)}(f>\fp)\propto f^{-2}$ for the gauge-field-induced one.

For BP2, with $\nir=1.0$ and $\nuv=-4.0$, the infrared behavior remains nearly unchanged, and the spectra therefore almost overlap for $f<\fp$. The steeper UV branch, however, leads to markedly different high-frequency scalings: $\ogw^{(\zeta)}\propto f^{-9.0}$ for the scalar-induced contribution and $\ogw^{(B)}(f>\fp)\propto f^{-5.0}$ for the gauge-field-induced one. The bottom panel of Fig.~\ref{fig:sgw_bp} shows the corresponding local spectral index $\ngw$, whose asymptotic behavior agrees with the analytical expressions derived in the main text.

These results demonstrate that the spectral distinction is not tied to a particular benchmark choice. While the peak frequency and normalization depend on the primordial parameters, the different UV responses persist because the two GW channels are generated through distinct tensor-source kernels. The broadband spectral shape therefore provides information about the origin of the induced GWs beyond their peak frequency and amplitude.

\bibliographystyle{apsrev4-1}
\bibliography{references}

\end{document}